\documentclass{aa}
\usepackage{graphicx,natbib,hyperref}
\usepackage[varg]{txfonts}
\usepackage{color}

\newcommand{\Hm}{\langle B\rangle}
\newcommand{\Hz}{\langle B_z\rangle}

\newcommand{\Hav}{\Hm_{\rm av}}

\newcommand{\Hzrms}{\Hz_{\rm rms}}

\newcommand{\Prot}{P_{\rm rot}}

\newcommand{\Zeeman}{\Delta\lambda_{\rm Z}}
\newcommand{\ew}{W_\lambda}

\newcommand{\Feline}{Fe~{\sc ii}~$\lambda\,6149.2$}
\newcommand{\Ndline}{Nd~{\sc iii}~$\lambda\,6145$}

\bibpunct[ ]{(}{)}{;}{a}{}{,}

\begin{document}

\title{HD~965: An extremely peculiar A star\\
  with an extremely long rotation period\thanks{Based on observations
  collected at the European
  Southern Observatory under ESO Programmes 52.7-063, 53.7-028,
  55.E-0751, 57.E-0637, 61.E-0711, and
  72.D-0138; at the Canada-France-Hawaii Telescope (CFHT), which is
  operated by the National Research Council of Canada, the Institut
  National des Sciences de l'Univers of the Centre National de la
  Recherche Scientifique of France, and the University of Hawaii; and 
at the 6-m telescope BTA of the Special Astrophysical Observatory of
the Russian Academy of Sciences. Also based on data products from
observations made with ESO Telescopes at the La Silla Paranal
Observatory under programmes 
68.D-0254, 70.D-0470, 74.C-0102, 75.C-0234, 76.C-0073, 76.D-0169,
79.C-0170, 279.D-5059, and 80.C-0032.}} 

\author{G.~Mathys\inst{1} 
  \and I.~I.~Romanyuk\inst{2,3}
  \and S. Hubrig\inst{4}
\and D.~O.~Kudryavtsev\inst{2}
\and M.~Sch\"oller\inst{5}
\and E.~A.~Semenko\inst{2,6}
\and I.~A.~Yakunin\inst{2}}

\institute{European Southern Observatory,
  Alonso de Cordova 3107, Vitacura, Santiago, Chile\\\email{gmathys@eso.org}
\and
Special Astrophysical Observatory, Russian Academy of Sciences,
Nizhnii Arkhyz, 369167 Russia
\and
Institute of Astronomy of the Russian Academy of Sciences, 48 Pyatnitskaya St, 119017 Moscow, Russia
\and
Leibniz-Institut für Astrophysik Potsdam (AIP), An der Sternwarte 16, 14482 Potsdam, Germany
\and
European Southern Observatory, Karl-Schwarzschild-Str. 2, 85748
Garching bei M\u unchen, Germany
\and
National Astronomical Research Institute of Thailand, Chiangmai, 50180
Thailand}

\date{Received $\ldots$ / Accepted $\ldots$}

\titlerunning{The rotation period of HD~965}

\authorrunning{G. Mathys et al.}

\abstract
{One of the keys to understanding the origin of the Ap stars and their
  significance in the general context of stellar astrophysics is the
  consideration of the most extreme properties displayed by some of
  them. In that context, HD~965 is particularly interesting, as it
  combines some of the most pronounced chemical peculiarities with one
of the longest rotation periods known.} 
{We characterise the variations of the magnetic field of the Ap star
  HD~965 and derive constraints about its structure.} 
{We combine published measurements of the mean longitudinal field
  $\Hz$ of HD~965 with new determinations of this field moment from
  circular spectropolarimetry obtained at the 6-m telescope BTA of the
  Special Astrophysical  Observatory of the Russian Academy of
  Sciences. For the mean magnetic field modulus $\Hm$, 
  literature data 
  are complemented by the analysis of ESO archive spectra.}
{We present the first determination of the rotation period of
  HD~965, $\Prot=(16.5\pm0.5)$\,y. HD~965 is only the
  third Ap star with a period longer than 10 years for which magnetic
  field measurements have 
  been obtained over more than a full cycle. The variation curve of
  $\Hz$ is well approximated by a cosine wave. $\Hm$  does not
  show any significant variation. The observed behaviour of
    these field moments is well represented by a simple model
    consisting of the superposition of collinear dipole, quadrupole
    and octupole. The distribution of
neodymium over the surface of HD~965 is highly non-uniform. The
element appears concentrated around the magnetic poles, especially the
negative one.} 
{The shape of the longitudinal magnetic variation
  curve of HD~965 indicates that its magnetic field is essentially
  symmetric about an axis passing through the centre of the
  star. Overall, as far as its magnetic field is concerned,
  HD~965 appears similar to the bulk of the
  long-period Ap stars.} 

\keywords{Stars: individual: HD~965 --
Stars: chemically peculiar --
Stars: rotation --
Stars: magnetic field}

\maketitle

\section{Introduction}
\label{sec:intro}
\citet{1963AJ.....68..428S} appears to have been the first to notice the
peculiarity of the spectrum of HD~965 (= BD~$-$0~21), a star of
spectral type A8p SrEuCr according to
\citet{2009A&A...498..961R}. Very little attention was paid to this
star until it was found to have resolved magnetically split lines
\citep{1997A&AS..123..353M}. The subsequent effort that was undertaken
to study its magnetic field suggested the possible occurrence of the
core-wing anomaly (CWA) in the hydrogen Balmer lines, a suspicion
that was soon confirmed by means of a dedicated observation
\citep{2001A&A...367..939C}. This indicates that the structure of the
atmosphere of HD~965 cannot be represented by a normal A-star
  model atmosphere. It has also been reported that the magnetic field
  of HD~965 decreases with increasing
optical depth \citep{2004A&A...422L..51N}, but more observations would
be needed for full characterisation of its structure.
More generally, the spectrum of HD~965 is
extremely peculiar, bearing considerable resemblance to that of
Przybylzki's star (HD~101065), the Ap star with the most unusual
spectrum \citep{2002ASPC..279..365H}. Like Przybylski's star, HD~965
shows in its spectrum spectral lines that have been tentatively
attributed to promethium, an unstable element \citep{2004A&A...419.1087C}. 

The presence of the CWA in its spectrum
motivated a photometric search for rapid oscillations in
HD~965. Despite the high precision of the observations, no such
oscillations were detected \citep{2003A&A...398.1117K}. A subequent
attempt using high-cadence high-resolution spectroscopy did not yield
any detection either \citep{2005MNRAS.358.1100E}.

Besides its extreme and intriguing spectral features, HD~965 also has
the distinction of rotating exceptionally
slowly. \citet{2017A&A...601A..14M} reported that its rotation
period must exceed 13\,y. This puts it firmly among the group of
Ap stars with rotation periods longer than one month. This group,
which has been progressively emerging in recent years, comprises
several percent of the whole Ap star class. Studying it is essential
for the understanding of the processes responsible for the
differentiation of stellar rotation rates. This has been discussed in
more detail elsewhere \citep{2019A&A...624A..32M}.

In this paper, we present new determinations of the mean longitudinal
magnetic field $\Hz$ and of the mean magnetic field modulus $\Hm$ of
HD~965. They are based on both new dedicated observations and archive
spectra. We use the complete set of existing $\Hz$ measurements of the
star to determine for the first time the value of its rotation
period. The observational data and their analysis are presented in
Sect.~\ref{sec:obs}, and the determination of the stellar rotation
period is described in Sect.~\ref{sec:per}. In
Sect.~\ref{sec:magfield}, we derive constraints on the geometrical
structure of the magnetic field. Finally, we discuss the implications
of the obtained results for HD~965 itself, and how it fits within the
general context of the slowly rotating Ap stars.

\section{Observations and data analysis}
\label{sec:obs}
\subsection{Mean magnetic field modulus}
\label{sec:Hm}
The mean magnetic field modulus $\Hm$ is the average over the visible
stellar hemisphere of the modulus of the field vector, weighted by the
local emergent line intensity.

All the mean field modulus values used in this analysis were
determined from the 
measured wavelength separation of the two magnetically split
components of the \Feline\ diagnostic line. The following formula was
applied to derive $\Hm$:
\begin{equation}
\lambda_{\rm r}-\lambda_{\rm b}=g\,\Zeeman\,\Hm\,.
\label{eq:Hm}
\end{equation}
In this equation, $\lambda_{\rm r}$ and $\lambda_{\rm b}$ are,
respectively, the wavelengths of the red and blue split line
components; $g$ is the Land\'e factor of the split level of the
transition ($g=2.70$; \citealt{1985aeli.book.....S}); 
$\Zeeman=k\,\lambda_0^2$, with
$k=4.67\,10^{-13}$\,\AA$^{-1}$\,G$^{-1}$; $\lambda_0=6149.258$\,\AA\
is the nominal wavelength of the considered transition.

The application of the above-described procedure is complicated by the
fact that the \Feline\ line is strongly blended on the blue side with
an unidentified rare earth line. This blend is present at different
degrees in a fraction of the Ap stars. As can be seen e.g. in
Figs.~2 to 4 of \citet{1997A&AS..123..353M}, HD~965 is one of the
stars in which its impact is most severe. Dealing with this blend as
consistently as possible in measuring the separation of the split
components of the \Feline\ line in spectra of HD~965 obtained at
different epochs is essential to achieve the best precision and
uniformity in the derived values of $\Hm$. This precision and
uniformity are required to ensure that the variations of $\Hm$ along
the stellar rotation cycle can be correctly characterised.

\begin{table}
\caption{Mean magnetic field modulus measurements.}
\label{tab:Hm}
\centering
\begin{tabular}{lrc}
\hline\hline\\[-4pt]
\multicolumn{1}{c}{JD}&\multicolumn{1}{c}{$\Hm$ (G)}&Configuration\\[4pt]
\hline\\[-4pt]
 2449301.597&4320&ESO CAT + CES SC\\
 2449534.896&4366&ESO CAT + CES LC\\
 2449535.879&4306&ESO CAT + CES LC\\
 2449881.918&4247&ESO CAT + CES LC\\
 2449908.843&4263&ESO CAT + CES LC\\
 2449947.836&4268&ESO CAT + CES LC (lower res.)\\
 2450231.914&4199&ESO CAT + CES LC\\
 2450296.576&4386&OHP 1.52\,m + AURELIE\\
 2450971.914&4256&ESO CAT + CES LC\\
 2451042.807&4326&ESO CAT + CES LC\\
 2451084.731&4192&ESO CAT + CES LC\\
 2451740.105&4234&CFHT + Gecko\\
 2451741.060&4264&CFHT + Gecko\\
 2452190.627&4303&ESO VLT UT2 + UVES\\
 2452420.063&4228&CFHT + Gecko\\
 2452535.723&4352&ESO VLT UT2 + UVES\\
 2452924.154&4293&ESO VLT UT2 + UVES\\
 2453215.093&4287&CFHT + Gecko\\
 2453334.505&4213&ESO 3.6\,m + HARPS\\
 2453581.741&4165&ESO 3.6\,m + HARPS\\
 2453582.764&4225&ESO 3.6\,m + HARPS\\
 2453583.867&4242&ESO 3.6\,m + HARPS\\
 2453661.702&4279&ESO VLT UT2 + UVES\\
 2453711.565&4237&ESO 3.6\,m + HARPS\\
 2453712.595&4177&ESO 3.6\,m + HARPS\\
 2453713.585&4201&ESO 3.6\,m + HARPS\\
 2453714.583&4238&ESO 3.6\,m + HARPS\\
 2453715.568&4208&ESO 3.6\,m + HARPS\\
 2453716.567&4245&ESO 3.6\,m + HARPS\\
 2454336.860&4185&ESO 3.6\,m + HARPS\\
 2454338.796&4215&ESO 3.6\,m + HARPS\\
 2454406.585&4310&ESO VLT UT2 + UVES\\
 2454442.613&4224&ESO 3.6\,m + HARPS\\
 2454443.560&4186&ESO 3.6\,m + HARPS\\
 2454469.580&4259&ESO VLT UT2 + UVES\\[4pt]
\hline
\end{tabular}
\end{table}

\begin{table*}
\caption{Mean longitudinal magnetic field measurements.}
\label{tab:Hz}
\begin{tabular*}{\textwidth}[]{@{}@{\extracolsep{\fill}}cc
@{\extracolsep{0pt}}@{}}
\parbox[t]{8.8cm}{
\centering
\begin{tabular}[t]{lrrrl}
\hline\hline\\[-4pt]
  \multicolumn{1}{c}{JD}&\multicolumn{1}{c}{S/N}&\multicolumn{1}{c}{$\Hz$}&\multicolumn{1}{c}{$\sigma_z$}&Reference\\
  &&\multicolumn{1}{c}{(G)}&\multicolumn{1}{c}{(G)}&\\[4pt]
\hline\\[-4pt]
2449916.907&140& $-$574&113&\protect{\citet{2017A&A...601A..14M}}\\
2449974.744&110& $-$691&141&\protect{\citet{2017A&A...601A..14M}}\\
2450039.588&120& $-$863& 44&\protect{\citet{2017A&A...601A..14M}}\\
2450294.905&120&$-$1063&125&\protect{\citet{2017A&A...601A..14M}}\\
2450615.928&120&$-$1057& 52&\protect{\citet{2017A&A...601A..14M}}\\
2450629.884&125&$-$1159&142&\protect{\citet{2017A&A...601A..14M}}\\
2450784.596&150&$-$1270&139&\protect{\citet{2017A&A...601A..14M}}\\
2451806.479&100& $-$430&120&\citet{2005MNRAS.358.1100E}\\
2452130.497&100& $-$360& 70&\citet{2005MNRAS.358.1100E}\\
2452153.356&40& $-$280&120&\citet{2005MNRAS.358.1100E}\\
2452625.209&90& $-$160& 80&\citet{2005MNRAS.358.1100E}\\
2452626.232&90&  $-$80& 50&\citet{2005MNRAS.358.1100E}\\
2452831.529&180&  $-$30& 80&\citet{2005MNRAS.358.1100E}\\
2453273.296&190&  320& 50&\citet{2005MNRAS.358.1100E}\\
2453362.231&300&  360& 30&\citet{2015AstBu..70..456R}\\
2453666.290&330&  330& 40&\citet{2015AstBu..70..456R}\\
2453667.270&250&  470& 40&\citet{2015AstBu..70..456R}\\
2453718.245&450&  420& 30&\citet{2015AstBu..70..456R}\\
2453953.425&350&  600& 40&\citet{2015AstBu..70..456R}\\
2454015.236&350&  420& 50&\citet{2015AstBu..70..456R}\\
2454402.285&300&  530& 50&\citet{2014AstBu..69..427R}\\
2455017.511&250&  150& 40&\citet{2016AstBu..71..302R}\\
2455075.438&300&  240& 50&\citet{2016AstBu..71..302R}\\
2455431.458&250&  $-$70& 20&\citet{2017AstBu..72..391R}\\
2455459.492&250& $-$240& 50&\citet{2017AstBu..72..391R}\\
2455461.462&300& $-$140& 50&\citet{2017AstBu..72..391R}\\
2455553.174&300& $-$340& 20&\citet{2017AstBu..72..391R}\\
2455555.139&200& $-$360& 30&\citet{2017AstBu..72..391R}\\
2455583.148&180& $-$359& 26&\citet{2018AstBu..73..178R}\\
2455842.411&250& $-$624& 23&\citet{2018AstBu..73..178R}\\
2455843.398&250& $-$643& 23&\citet{2018AstBu..73..178R}\\
2455871.176&150& $-$470& 29&\citet{2018AstBu..73..178R}\\
2455962.138&150& $-$590& 30&\citet{2015AstBu..70..456R}\\[4pt]
\hline
\end{tabular}}
&\parbox[t]{8.8cm}{
\centering 
\begin{tabular}[t]{lrrrl}
\hline\hline\\[-4pt]
  \multicolumn{1}{c}{JD}&\multicolumn{1}{c}{S/N}&\multicolumn{1}{c}{$\Hz$}&\multicolumn{1}{c}{$\sigma_z$}&Reference\\
  &&\multicolumn{1}{c}{(G)}&\multicolumn{1}{c}{(G)}&\\[4pt]
\hline\\[-4pt]
2456174.540&200& $-$670& 40&\citet{2015AstBu..70..456R}\\
2456177.429&300& $-$800& 50&\citet{2015AstBu..70..456R}\\
2456234.208&110&$-$1030& 30&\citet{2015AstBu..70..456R}\\
2456500.509&160&$-$1190& 20&\citet{2015AstBu..70..456R}\\
2456589.383&160&$-$1250& 20&\citet{2015AstBu..70..456R}\\
2456640.125&150&$-$1130& 50&\citet{2015AstBu..70..456R}\\
2456940.302&150&$-$1170& 30&\citet{2015AstBu..70..456R}\\
2456967.292&150&$-$1130& 60&\citet{2015AstBu..70..456R}\\
2456995.144&200&$-$1360& 50&\citet{2015AstBu..70..456R}\\
2457169.525&150&$-$1340& 30&\citet{2015AstBu..70..456R}\\
2457246.465&200&$-$1200& 20&\citet{2015AstBu..70..456R}\\
2457288.469&200&$-$1170& 20&This paper\\
2457352.232&150&$-$1110& 40&This paper\\
2457355.195&200&$-$1120& 20&This paper\\
2457592.479&120&$-$1090& 40&This paper\\
2457740.260&100& $-$910& 60&This paper\\
2457761.132&130& $-$960& 30&This paper\\
2457762.156&120& $-$920& 30&This paper\\
2457764.149&120&$-$1070& 50&This paper\\
2457950.436&180& $-$990& 30&This paper\\
2458006.458&110& $-$780& 20&This paper\\
2458009.364&100& $-$820& 30&This paper\\
2458061.227&130& $-$680& 30&This paper\\
2458067.216&150& $-$780& 30&This paper\\
2458116.161&140& $-$610& 30&This paper\\
2458117.213&150& $-$690& 30&This paper\\
2458126.137&150& $-$650& 20&This paper\\
2458151.136&120& $-$740& 20&This paper\\
2458447.277&150& $-$320& 20&This paper\\
2458448.298&170& $-$340& 30&This paper\\
2458449.162&100& $-$290& 40&This paper\\
2458512.155&160& $-$260& 40&This paper\\
  \\[4pt]
\hline
\end{tabular}}\\
\end{tabular*}
\end{table*}

Obviously, we cannot be sure that the way in which we measure
HARPS and UVES archive spectra at present is consistent with the way
in which spectra obtained (mostly) with other instruments were
measured in the past to derive the published values of $\Hm$
\citep{1997A&AS..123..353M,2005MNRAS.358.1100E,2017A&A...601A..14M}.
Accordingly, to ensure that the set of mean magnetic field modulus
data under consideration is as uniform as possible, we remeasured
all the spectra of the cited references and used them to derive
updated $\Hm$ values. Thus, the full set of observations that we
analysed consists of:
\begin{itemize}
  \item 9 spectra recorded with the Long Camera (LC) of the Coud\'e
    Echelle Spectrograph (CES) fed by the ESO Coud\'e Auxiliary
    Telescope (CAT), one of which had a somewhat lower resolution
    than the others owing to a detector problem;
    \item 1 spectrum recorded with the Short Camera (SC) of the CES,
      fed by the CAT;
      \item 1 spectrum recorded with the AURELIE spectrograph fed by
        the 1.52-m telescope of the Observatoire de Haute-Provence
        (OHP);
        \item 4 spectra recorded with the Gecko spectrograph fed by
          the Canada-France-Hawaii Telescope (CFHT);
          \item 6 spectra recorded with the Ultraviolet and Visible
        Echelle Spectrograph (UVES) fed by Unit Telescope 2 (UT2) of
        the ESO Very Large Telescope (VLT). One of them is the average
        of the 111 high-cadence spectra analysed by
        \citet{2005MNRAS.358.1100E}; the others were retrieved from
        the ESO Archive;
           \item 14 spectra recorded with the High Accuracy Radial velocity
      Planet Searcher (HARPS) fed by the ESO 3.6-m telescope,
      retrieved from the  ESO Archive.
    \end{itemize}
    More details about the first four configurations of this list and
    the corresponding data reduction processes are provided by
    \citet{1997A&AS..123..353M}. See also \citet{2005MNRAS.358.1100E}
    for information about the reduction of the respective
    observations. For the other UVES observations, and for the HARPS
    spectra, we used science grade pipeline
processed data available from the ESO Archive. The only additional
processing that we carried out was a continuum normalisation of the
region ($\sim$100\,\AA\ wide) surrounding the \Feline\ diagnostic line. 

The challenge that we met to achieve consistent measurements of the
wavelength separation of the split components of the \Feline\ line
despite the heavy blending of its blue wing led us to revise the
estimate of the uncertainty affecting the derived $\Hm$ values that
had been adopted by \citet{1997A&AS..123..353M} and by
\citet{2017A&A...601A..14M}. The latter, 30\,G, is definitely too
optimistic, as it is of the same order as the estimated $\Hm$
uncertainty for stars such as HD~50169, for which measuring the
wavelength splitting of the \Feline\ line is definitely much more
straightforward \citep{2019A&A...624A..32M}. Taking this into
account, and after comparing the results of measurements of a number
of spectra repeated multiple
times, we now adopt 50\,G as a revised estimate, hopefully more
realistic, of the uncertainty affecting the determinations of the mean
magnetic field modulus of HD~965. 

The 35 values of the mean magnetic field modulus obtained in the
above-described way are
presented in Table~\ref{tab:Hm}. The
columns give, in order, the Heliocentric (or Barycentric, for HARPS)
Julian Date of mid-exposure, the value
$\Hm$ of the mean magnetic field modulus, and the instrumental
configuration with which the analysed spectrum was obtained.

\subsection{Mean longitudinal magnetic field}
\label{sec:Hz}
The mean longitudinal magnetic field $\Hz$ is the average over the
visible hemisphere of the component of the magnetic vector along the
line of sight, weighted by the local emergent line intensity. The
following published measurements of this field moment were used in
this study: 
\begin{itemize}
\item 7 measurements from \citet{2005MNRAS.358.1100E};
\item 1 measurement from \citet{2014AstBu..69..427R};
\item 18 measurements from \citet{2015AstBu..70..456R};
\item 2 measurements from \citet{2016AstBu..71..302R};
\item 7 measurements from \citet{2017A&A...601A..14M};
\item 5 measurements from \citet{2017AstBu..72..391R};
\item and 4 measurements from \citet{2018AstBu..73..178R}.
\end{itemize}

All these measurements were carried out through the analysis of medium
spectral resolution observations of metal lines in circular
polarisation. The data from \citet{2017A&A...601A..14M} are based on
observations obtained with the CASPEC spectrograph mounted on the ESO
3.6\,m telescope, in the post-1995 configuration specified by
\citet{1997A&AS..124..475M}. All other listed studies were carried out
with the instrumentation introduced in the next paragraph. 

Here the measurements from the literature are complemented by additional $\Hz$
determinations obtained from observations of the same type, that is,
spectra of HD~965 recorded at $R\simeq14\,500$ in both circular
polarisations with the Main Stellar Spectrograph of the 6-m telescope
BTA of the Special Astrophysical Observatory (SAO), on 21 nights spread from
September 2015 to January 2019. This is the same configuration as used
by \citet{2014AstBu..69..427R}, and in the other works of the same
group. The instrumental configuration and the data 
reduction procedure are as described in detail in this reference.

The mean longitudinal magnetic field was determined from the
wavelength shifts of a sample of spectral lines between 
the two circular polarisations in each of these spectra, by application of the formula:
\begin{equation}
\lambda_{\rm R}-\lambda_{\rm L}=2\,\bar g\,\Zeeman\,\Hz\,,
\label{eq:Hz}
\end{equation}
where $\lambda_{\rm R}$ (resp. $\lambda_{\rm L}$) is the wavelength of
the centre of gravity of the line in right (resp. left) circular
polarisation and $\bar g$ is the effective Land\'e factor of the
transition. $\Hz$ is determined through a 
least-squares fit of the measured values of $\lambda_{\rm
  R}-\lambda_{\rm L}$ by a function of the form given above. The standard error
$\sigma_z$ that is derived from that
least-squares analysis is used as an estimate of the uncertainty
affecting the obtained value of $\Hz$. The implementation details
have been described by 
\citet{2017A&A...601A..14M} for the analysis of the CASPEC observations,
and by \citet{2014AstBu..69..427R} and by \citet{2015AstBu..70..456R} for the SAO
spectra. Whenever both 
the modernised Babcock method and the regression method were applied for
determination of $\Hz$ from these spectra, the value yielded by the
regression method was adopted. For HD~965, the
differences between the mean longitudinal values derived through
application of either method \citep{2015AstBu..70..456R} are insignificant.


The values of the mean longitudinal field obtained in the way
described above are
presented in Table~\ref{tab:Hz}. For the convenience of the reader,
this table also includes the previously published measurements. In
total, 65 $\Hz$ measurements are analysed in this study. The table
columns give, in order, the Heliocentric Julian Date of
mid-observation, the mean S/N ratio of the recorded spectrum, the value
$\Hz$ of the mean longitudinal magnetic field and its uncertainty
$\sigma_z$, and the source of the measurement. 

\section{Variability and rotation period}
\label{sec:per}
To determine the rotation  period of HD~965, we fitted the
measurements of its mean longitudinal magnetic field by either a cosine
wave, or the superposition of a cosine wave and of its first
harmonic, progressively varying the period of these waves, in search
of the value of the period that minimises the reduced $\chi^2$ of the
fit. These fits are weighted by the inverse of the square of
  the uncertainties of the individual measurements. Through this
  procedure, we concluded unambiguously that the 
rotation period of the star must be of the order of 6000\,d. The
corresponding periodogram is shown in Fig.~\ref{fig:periodo}. 

The time elapsed between the first determination of the mean
longitudinal magnetic field of HD~965 by \citet{2017A&A...601A..14M}
and our most recent 
spectropolarimetric observation of the star is 8596\,d, or $\sim$1.4
rotation period. This provides a very sensitive approach to refine
the determination of the period and to estimate its uncertainty. By
plotting a phase diagram of the $\Hz$ measurements for a series of
tentative values of the period around the one suggested by the
periodogram, one can visually identify the period value that minimises
the phase shifts between field determinations from different rotation
cycles, and constrain the range around that value for which those
phase shifts remain reasonably small.

 Determinations of $\Hz$ obtained with different
instruments and by application of different data analysis methods are
known to show frequently systematic differences, whose existence
between the two sets of data of interest cannot be definitely ruled
out. However, previous studies indicate that, in general, for Ap stars
with fairly sharp spectral lines, there are at most minor
systematic differences between the longitudinal field values
determined, through application of the metal line spectropolarimetric
technique, by Mathys
and collaborators (from ESO CASPEC spectra), and by  Romanyuk and
collaborators (with the Main Stellar Spectrograph of the 6-m telescope
BTA of the SAO). One can also note that, as the spectral range spanned
by the SAO spectra has increased over time
\citep{2015AstBu..70..456R}, different sets of lines have been
measured at different epochs, without introducing any obvious
inconsistency in the derived $\Hz$ values. 

Assuming that the $\Hz$
datasets that we combine here are indeed mutually consistent, we
derived the following best value of the rotation period of HD~965:
\begin{equation}
  \Prot=(6030\pm200)\,\mathrm{d}.
  \label{eq:Prot}
  \end{equation}

\begin{figure}
\resizebox{\hsize}{!}{\includegraphics{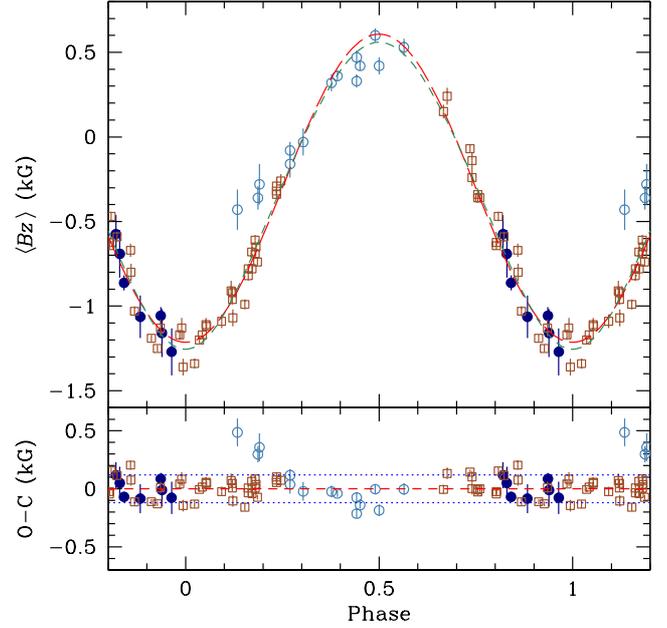}}
\caption{\textit{Upper panel:\/} Mean longitudinal magnetic field of
  HD~965 against rotation 
  phase. To distinguish measurements obtained at similar phases in
  different cycles, circles (blue) were used to identify observations
  acquired until 2008; squares (brown) correspond to spectra taken since
  2009. Filled symbols are used for the measurements of
  \citet{2017A&A...601A..14M}, and open symbols for all determinations
  based on SAO spectra. The
long-dashed line (red) is the best fit of the observations by a cosine wave
 -- see Eq.~(\ref{eq:bzfit}).
The short-dashed
line (green) corresponds to the superposition of low-order multipoles
discussed in Sect.~\ref{sec:magfield}.
\textit{Lower panel:\/}
Differences $\mathrm{O}-\mathrm{C}$ between the individual $\Hz$
measurements and the best fit curve, against rotation phase. The
dotted lines (blue) correspond to $\pm1$~rms deviation of the
observational data about the fit (red dashed line). The symbols are
the same as in the upper panel.} 
\label{fig:bzcurve}
\end{figure}

 The phase variation curve of the mean longitudinal magnetic field for
this value of the period is 
shown in Fig.~\ref{fig:bzcurve}. Note the consistent behaviour of the
measurements of \citet{2017A&A...601A..14M} and of Romanyuk's group,
on the descending 
branch between phases 0.82 and 0.96. The apparent inconsistency of the
first three measurements of \citet{2005MNRAS.358.1100E} (open circles
at phases 0.13--0.19) with the rest of the data likely reflects the
fact that some early SAO measurements are affected by stochastic
errors greater than indicated by the formal uncertainties. These
errors are related to instrumental shifts that could not always be fully
corrected through the observation of null standards. In support of this
interpretation of the observed inconsistency, we note that the
increase in the length of the adopted 
period that would be required to make the three $\Hz$ values
under consideration consistent with the 2018--2019 SAO measurements
would imply 
an irreconcilable discrepancy between the ESO and SAO data, since the
former would still show a monotonic decrease at phases at which the
latter have already started to increase from the negative $\Hz$
extremum towards 0. There is no plausible way to account for such a
different behaviour by systematic effects between different
instruments.  

The application of the above-described period determination procedure
has been justified in detail by \citet{2019A&A...624A..32M}. It must
be stressed 
that the accuracy of the period determination
  carried out in this manner only depends on the reproducibility of
  the variation curve from one cycle to the next, not on its exact
  shape. The adopted uncertainty of the derived period value is based
  on the assumption that any systematic offset between the mean
  longitudinal magnetic field measurements obtained with the two
  different telescope-instrument combinations does not significantly
  exceed the random errors affecting these measurements. For a more
  detailed discussion of this point, see \citet{2019A&A...624A..32M}.

\begin{figure}
\resizebox{\hsize}{!}{\includegraphics[angle=270]{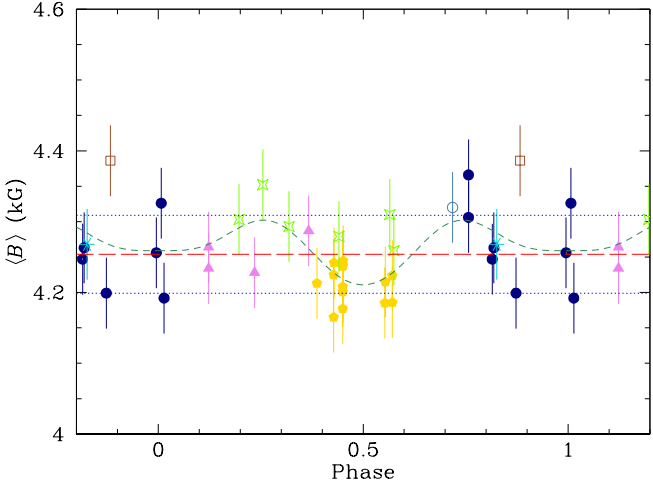}}
\caption{Mean
  magnetic field modulus of HD~965 against rotation
  phase. The different symbols identify the instrumental
  configuration from which the $\Hm$ value was obtained, as follows:
  filled circles (dark blue): CAT + CES LC; asterisk (turquoise): CAT
  + CES LC, lower resolution; open circle (steel blue):
  CAT + CES SC; open square (brown): OHP 1.52\,m + AURELIE; filled
  triangles (violet): CFHT + Gecko \citep[all previous symbols
  identical to][]{2017A&A...601A..14M}; four-pointed open stars (light
  greeen): 
  UT2 + UVES; filled pentagons (yellow): 3.6\,m + HARPS.
The short-dashed
line (green) corresponds to the superposition of low-order multipoles
discussed in Sect.~\ref{sec:magfield}.
The
dotted lines (blue) correspond to $\pm1$~standard deviation of the
individual measurements about their average value (red dashed line).}
\label{fig:bmcurve}
\end{figure}

 The measurements of the mean magnetic field modulus $\Hm$ are also
plotted against rotation phase, in Fig.~\ref{fig:bmcurve}. There is no
apparent trend; the $\Hm$ values do not show any dependence on
the observation phase. Furthermore, the scatter of
these values is consistent with their estimated uncertainties: the
average of our 34 measurements is $\Hav=4254$\,G, with a standard
deviation of 55\,G. This
leads us to conclude that the mean field modulus of HD~965 is constant
at the achieved precision level.

On the other hand, consideration of Fig.~\ref{fig:bmcurve} suggests that the
$\Hm$ measurements obtained from HARPS spectra may be slightly but
systematically shifted
downward, and those obtained from UVES spectra, shifted
upward, with respect to the values obtained with the
CES and Gecko. This may be indicative of
some amount of systematic error of instrumental origin. However, the
magnitude of the observed offset, a few tens of G, is small, at most of
the same order as the random measurement uncertainties, so that one
should be careful not to overinterpret it. It is also noteworthy
that the highest $\Hm$ value was obtained from the only AURELIE
observation. This again may arise from an instrumental effect, since
AURELIE is known to
be definitely subject to such effects
\citep{1997A&AS..123..353M}. However, no 
definitive conclusion can be drawn from consideration of a single
measurement point.

In any event, whether the instrumental effects discussed above are
actually present or not, their magnitude is small enough so that they
do not question the conclusion that the mean magnetic field modulus of
HD~965 does not show any significant variability. 

One observable that shows prominent variability with the same
period as $\Hz$ is the intensity of the \Ndline\ line. While the
high-resolution spectra from which we determined $\Hm$ cover a variety
of spectral ranges, some of them very narrow, the \Ndline\ line is
present in all of them. The values of the equivalent width $\ew$ of
this line that we measured from these high-resolution spectra are
plotted against the rotation phase in Fig.~\ref{fig:ew}.\footnote{The
  equivalent width was not measured in the single AURELIE spectrum,
  since experience indicates that equivalent widths determinations
  with this instrument may be affected by significant errors owing to
  uncertainties in the contribution of the dark current of the
  detector in spectra of faint stars such as HD~965
  \citep[e.g.][]{1997A&AS..123..353M}.} 
  They follow a
very definite double-wave variation curve, with a primary maximum
close to the phase of the negative extremum of the mean longitudinal
magnetic field, and a secondary maximum near the phase of the positive
extremum of $\Hz$. The observed variation of the equivalent width of
the \Ndline\ line in HD~965 represents an independent confirmation of
the rotation period of this star.

As noted by \citet{2017A&A...601A..14M}, the equivalent widths of the
Fe lines do not show any variability, so that there is no
indication of departures from uniformity in the distribution of this
element over the stellar surface. The
  equivalent width and magnetic resolution of several of the Fe lines
  are of the same order as for the
  \Ndline\ line: this represents a strong indication that the observed
  variations of the equivalent width of the latter reflect the
  inhomogeneous distribution of Nd over the stellar surface, not the
  occurrence of non-uniform magnetic desaturation.

\begin{figure}
\resizebox{\hsize}{!}{\includegraphics[angle=270]{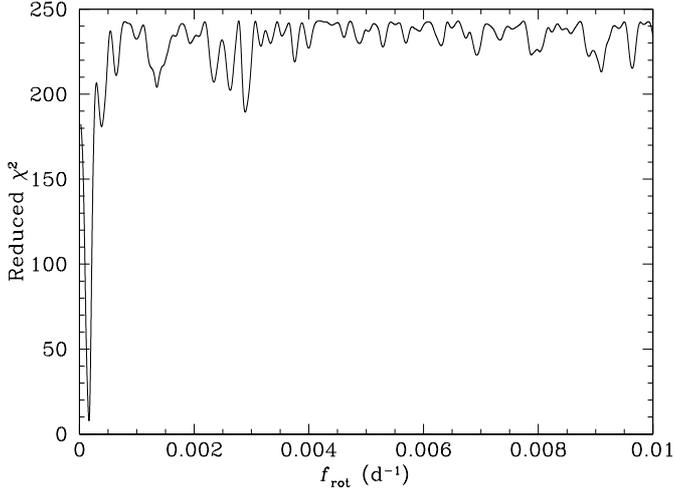}}
\caption{Periodogram of the variations of the mean longitudinal magnetic field
  of HD~965. The ordinate is the reduced $\chi^2$ of a fit of
  the $\Hz$ measurements by a cosine wave, with the frequency given in
  abscissa. }
\label{fig:periodo}
\end{figure}
 
\section{Magnetic field characterisation}
\label{sec:magfield}
The observed variations of the mean longitudinal magnetic field
of HD~965 can be well
represented by a cosine wave. The best least-squares fit solution for $\Prot=6030$\,d
is:
\begin{eqnarray}
\Hz(\phi)&=&(-306\pm14)\nonumber\\
&+&(911\pm21)\,\cos\{2\pi\,[\phi-(0.501\pm0.003)]\}\nonumber\\
         &&\hspace{-3em}(\nu=62,\ \chi^2/\nu=7.9),\label{eq:bzfit}
\end{eqnarray}
where the field strength is expressed in gauss, $\phi=({\rm
  HJD}-{\rm HJD}_0)/\Prot$ (mod 1) and the adopted 
value of ${\rm HJD}_0=2451000.0$ corresponds within the uncertainty of
the phase to a negative extremum of
the mean longitudinal magnetic field; $\nu$ is the number of degrees of
  freedom, and $\chi^2/\nu$, the reduced $\chi^2$ of the fit. The fitted curve
is shown in Fig.~\ref{fig:bzcurve}; the $\mathrm{O}-\mathrm{C}$
differences between the individual 
  measurements and this curve are also illustrated. The rather high
  value of $\chi^2/\nu$ may indicate that the uncertainty of
  some of the measurements is slightly underestimated, that there are
  some (small) systematic differences between the measurements of the
  different groups, or that the actual shape of the $\Hz$ variation
  curve departs somewhat from a cosine wave -- or some combination of
  these effects. However, no mathematically significant contribution
  of the first harmonic was found. In particular, the $\Hz$ variation
  curve appears to be very nearly mirror-symmetric about the phases of
  its extrema. 
  
\begin{figure}
\resizebox{\hsize}{!}{\includegraphics[angle=270]{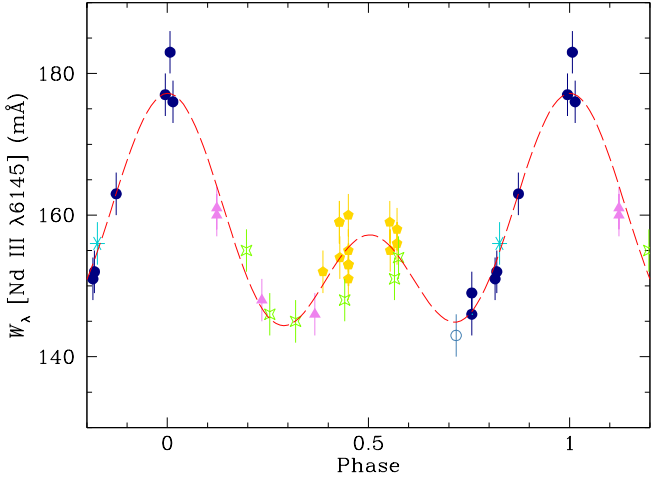}}
\caption{Equivalent width of the \Ndline\ line against rotation
  phase. The symbols have the same meaning as in
  Fig.~\ref{fig:bmcurve}. The
long-dashed line (red) is the best fit of the measurements by a
cosine wave and its first harmonic. It has been included to guide the
eye despite its lack of physical meaning. 
}
\label{fig:ew}
\end{figure}

The reversal of the sign of the mean longitudinal field over
  the rotation cycle indicates that both magnetic poles of the star
  come alternatively into sight.
  The constancy of the mean magnetic field modulus in a star of which
both magnetic poles come alternatively into sight is unique
\citep[see Fig.~9 of][]{2017A&A...601A..14M}. Combined with the
large amplitude and the sign reversal of the mean longtiudinal
field, it sets a strong constraint on the geometrical structure
of the magnetic field of HD~965.

In contrast with other extremely slowly rotating Ap stars that were
recently studied, such as HD~18078 \citep{2016A&A...586A..85M} or
HD~50169 \citep{2019A&A...624A..32M}, neither the mean longitudinal
magnetic field nor the mean magnetic field modulus of HD~965 show any
hint of significant deviations of the geometrical structure of the large-scale
magnetic field from axisymmetry. Accordingly, a simple axisymmetric model such
as the superposition of collinear dipole, quadrupole, and octupole
\citep{2000A&A...359..213L} may in principle be adequate to represent
this structure. We computed such a model using a Python programme that
is based on J.~D.~Landstreet's {\sc Fldcurv} code
\citep[e.g.,][]{2006A&A...451..195W} and solves a non-linear
least-squares minimisation problem,  fitting both the $\Hz$ and $\Hm$
variation curves simultaneously. The best parameter values are
$i=75.4^\circ\pm0.8^\circ$, $\beta=37.2^\circ\pm1.1^\circ$, $B_{\rm
  dipole}=(-5653\pm119)$\,G. $B_{\rm
  quadrupole}=(-281\pm171)$\,G. and $B_{\rm
  octupole}=(2872\pm377)$\,G.  As usual, the
angles $i$ (inclination of the rotation axis on the line of sight) and
$\beta$ (angle between the rotation and magnetic axes) can be
interchanged with no changes in the predicted variation curves. These
curves are shown as dark green, short-dashed lines in
Figs.~\ref{fig:bzcurve} and \ref{fig:bmcurve}. One should note the
similarity of the model curve to the cosine best fit curve for the
mean longitudinal magnetic field, and the fact that the amplitude of
the model curve for the mean magnetic field modulus is lower than the
standard deviation of the individual meaurements of this field moment
about their mean. The quadrupolar component of the model is almost
negligible. The two main components, a dipole and an octupole with 
opposite orientations, give a good approximation of a field that, in
$\Hz$, appears essentially dipolar, but for which $\Hm$ indicates that
the modulus of the field vector is nearly constant over the entire
visible part of the stellar surface. However, while the
  adopted model provides a satisfactory representation of the observed
  variations, one cannot be sure that it is physically meaningful, and
  accordingly, 
  the uncertainties given for the model parameters are only formal.

\section{Discussion}
\label{sec:concl} 
With a rotation period of 6030\,d, or 16.5\,yr, HD~965 is only the
third Ap star with a rotation period longer than 10\,yr for which the
value of this period has been determined exactly, and well sampled
magnetic variation curves have been obtained. 

Despite the extreme character of HD~965 in several respects (see
Sect.~\ref{sec:intro}), its magnetic field does not appear
extraordinary in any way. Its most remarkable property is the lack
of significant variability of $\Hm$ within the excellent precision of
the measurements of this field moment (within 50\,G). While large
values of the ratio $q$ between the extrema of the mean magnetic field
modulus are more frequently observed in slowly rotating Ap stars than
in those that rotate faster \citep[see Fig.~5
of][]{2017A&A...601A..14M}, this ratio is small in some of the
long-period stars. HD~965 is becoming the longest-period star known
to have $q\sim1$. This does not make it exceptional in that respect,
but rather results from the fact that among the slowly rotating Ap stars
whose periods have been determined exactly, only two have longer
periods. Once longer periods are determined for more stars, some of
them will likely show low amplitude variability of $\Hm$. 

The combination of $q\approx1.0$ with a value $r=-0.50$ of the ratio
of the smaller (in absolute value) to the larger (in absolute value)
extremum of $\Hz$ is somewhat more remarkable, since large relative
amplitudes of the mean magnetic field modulus tend to be found in
stars for which both poles come alternately into view \citep[see Fig.~9
of][]{2017A&A...601A..14M}. But it should not be regarded as a major
anomaly; other stars in which both magnetic poles are seen show $\Hm$
variations of low relative amplitude.

The nearly constant value of the mean magnetic field modulus of
HD~965, which 
averages at $\Hav=4254$\,G, is well below the value of 7.5\,kG that
seems to represent an upper limit to the field strength for those Ap
stars with rotation periods longer than 150\,d
\citep{1997A&AS..123..353M,2017A&A...601A..14M}. 
With a value of 0.18, the ratio of the
rms longitudinal magnetic field \citep{1993A&A...269..355B},
$\Hzrms=776$\,G, to $\Hav$, is also well within the typical range
\citep[see Fig.~8 of][]{2017A&A...601A..14M}.

Even though the modulus of the magnetic field vector must be mostly
constant over the surface of HD~965, the distribution of the abundances
of at least some elements (in particular, Nd)  is markedly
inhomogeneous. This suggests, perhaps not surprisingly,
that the local orientation of the magnetic
field must be more of a determining factor than its strength in the
formation of chemical spots on the surfaces of Ap stars.

\citet{2005MNRAS.358.1100E} suggested that the null results of the two
attempts that were made to detect pulsations in HD~965 could possibly be
attributed to the mostly equator-on aspect of the star at the epochs
when the observations were performed. The phases of these observations
can now be determined: 0.249 for the photometric study of
\citet{2003A&A...398.1117K}, and 0.319 for the high-resolution
spectroscopic recordings of
\citet{2005MNRAS.358.1100E}. Consideration of Fig.~\ref{fig:bzcurve}
confirms that both observations sampled primarily the part of the
stellar surface close to the magnetic equator. Accordingly, it may be
worth making a new attempt at detecting non-radial pulsations in
HD~965 close to a magnetic extremum, when one of the magnetic poles,
around which the pulsation amplitude is expected to be highest, is at
maximum visibility. The next opportunity to do so will arise close to
the forthcoming positive extremum of the mean longitudinal magnetic field, in
the second half of 2022.

\begin{acknowledgements}
Parts of this study were carried out during
a stay of GM in the Department of Physics \& Astronomy of the
University of Western Ontario (London, Ontario, Canada) funded by the
ESO Science Support Discretionary Fund (SSDF), and a stay of SH at the
ESO office in Santiago within the framework of the ESO Santiago
visitors programme. Thanks are due to ESO for its financial support of
these science stays, and to the respective host institutions for their
welcome. IIR (RSF grant No. 18-12-00423) gratefully acknowledges
the Russian Science Foundation for partial financial support.
This research has made use of the SIMBAD
database, operated at the CDS, Strasbourg, France.
\end{acknowledgements}

\bibliographystyle{aa}
\bibliography{hd965per}
\end{document}